\documentclass[twocolumn,preprintnumbers,amsmath]{revtex4}
\usepackage{graphicx}
\usepackage{color}
\usepackage{dcolumn}
\usepackage{bm}
\usepackage{epsfig}
\usepackage{amssymb}

\begin{document}

\title{The yielding dynamics of a colloidal gel}

\author{Thomas Gibaud, Damien Frelat, S\'{e}bastien Manneville\footnote[2]{e-mail: sebastien.manneville@ens-lyon.fr}}

\affiliation{Laboratoire de Physique, Universit\'{e} de Lyon -- {\'E}cole Normale Sup\'{e}rieure de Lyon -- CNRS UMR 5672, 46 all\'{e}e d'Italie, 69364 Lyon cedex 07, France}

\date{\today}

\begin{abstract}
Attractive colloidal gels display a solid-to-fluid transition as shear stresses above the yield stress are applied. This shear-induced transition is involved in virtually any application of colloidal gels. It is also crucial for controlling material properties. Still, in spite of its ubiquity, the yielding transition is far from understood, mainly because rheological measurements are spatially averaged over the whole sample. Here, the instrumentation of creep and oscillatory shear experiments with high-frequency ultrasound opens new routes to observing the local dynamics of opaque attractive colloidal gels. The transition proceeds from the cell walls and heterogeneously fluidizes the whole sample with a characteristic time whose variations with applied stress suggest the existence of an energy barrier linked to the gelation process. The present results provide new test grounds for computer simulations and theoretical calculations in the attempt to better understand the yielding transition. The versatility of the technique should also allow extensive mesoscopic studies of rupture mechanisms in soft solids ranging from crystals to glassy materials.
\end{abstract}

\maketitle

\section{Introduction}

Colloidal particles with strong enough short-range attraction are known to aggregate into fractal clusters that may form a space-spanning network leading to solid-like properties even at very low volume fractions \cite{Russel:1989,Larson:1999}. Such gelation obviously has tremendous consequences in terms of viscosity increase and rheological properties, which are capital for applications ranging from paints and coatings to food products and cosmetics. In the gel state, the system is kinetically arrested and displays non-ergodic features such as ultraslow relaxations, aging, and dynamical heterogeneities \cite{Cipelletti:2000,Duri:2006,Trappe:2007}. These glassy-like features have raised the issue of unifying the behaviour of colloidal gels and that of more concentrated amorphous jammed systems, e.g. colloidal glasses or concentrated emulsions \cite{Trappe:2001}. Therefore, most recent works on systems of short-range attractive particles have focused on the physics of gelation and on their structural properties at rest \cite{Cardinaux:2007,Gao:2007,Lu:2008,Royall:2008,Zaccarelli:2009}.

Of even greater practical importance is the behaviour of such materials under external stress. It is well known that applying stress to a soft glassy material generally triggers a solid-to-fluid transition, in the same way as increasing the temperature $T$, lowering the interparticle attractive energy $U$, or lowering the particle volume fraction $\phi$ \cite{Trappe:2001,OHern:2003}. In particular, the yielding phenomenon, i.e. the fact that the material flows like a liquid above a yield stress $\sigma_y$ and remains solid below $\sigma_y$, has been extensively documented in colloidal gels and glasses through standard rheological measurements and theoretical or numerical models, with emphasis on the scaling of viscoelastic properties and of $\sigma_y$ with $\phi$ and $U$ \cite{Russel:1989,Trappe:2001,Woutersen:1990,Potanin:1995,Rueb:1997,Whittle:1998,Barnes:1999,Fuchs:2002,Kobelev:2005c,Osuji:2008,Brader:2009}. However, rheology provides information that are averaged over the whole sample size and it cannot capture local phenomena such as apparent wall slip \cite{Barnes:1995}, shear localization \cite{Coussot:2002a}, or fracture within the material \cite{Pignon:1996}. Recently, a few original experimental approaches have been reported using confocal microscopy \cite{Besseling:2007,Isa:2007,Ballesta:2008b} and particle imaging velocimetry in microchannels \cite{Goyon:2008}. Still, these works were restricted to steady-state measurements, and, as a matter of fact, very little is known on how the stress-induced solid-to-fluid transition exactly proceeds in both space and time.

In the present article, we introduce high-frequency ultrasound as a new tool to follow the local deformation and flow of carbon black suspensions, a totally opaque colloidal gel made of weakly attractive soot particles. Both creep tests and oscillatory shear experiments show for the first time that, above a critical yield stress $\sigma_y$, (i) yielding proceeds from the cell walls in a spatially and temporally heterogeneous way and (ii) the time needed for total fluidization of the sample decreases exponentially with the applied shear stress. These results point out the existence of an energy barrier that linearly decreases with the external stress. Therefore, they provide evidence for the relevance of activated barrier hopping and for processes reminiscent of heterogeneous nucleation, in strong support of recent models \cite{Bocquet:2009,Fielding:2009}. Such features should be common to all weakly attractive colloids and models are expected to become quantitatively predictive, based on more spatially and temporally resolved measurements as described here.

\section{Materials and methods}

\subsection{Carbon black gel preparation}
Carbon black is widely used in industry, e.g. for coatings, printing, rubbers, tires, paints, or batteries \cite{Donnet:1993}. These colloidal carbon particles result from partial combustion of fuel and are made of unbreakable aggregates of permanently fused nanometric primary particles. These aggregates have a typical diameter of 500~nm. When dispersed in a light mineral oil (from Sigma, density 0.838, viscosity 20~mPa.s) at a weight concentration of a few percents, carbon black particles (Cabot Vulcan XC72R of density 1.8) form a space-spanning network of fractal dimension $d_f=2.2$ due to weakly attractive interactions of typical strength $U\sim 30 k_B T$ \cite{Trappe:2007}. Our samples are prepared by vigorous mixing of 6~\% w/w carbon black particles into the mineral oil. Since the resulting gel is transparent to ultrasound, 1~\% w/w hollow glass microspheres of mean diameter 6~$\mu$m (Sphericel, Potters) are added in order to provide enough acoustic contrast (see below). The sample is further sonicated for 1~hour.

\subsection{Standard rheology}
Rheological measurements are performed in a Plexiglas Couette cell (rotating inner cylinder radius $24$~mm, gap width 1~mm, and height 30~mm) by a stress-imposed rheometer (Bohlin C-VOR 150, Malvern Instruments). To ensure an initial reproducible gel state, all measurements are carried out by first preshearing the suspension at 1000~s$^{-1}$ then at $-1000$~s$^{-1}$ for 20~s each, in order to break up any large aggregate. The sample is then left at rest for 100~s during which the gel structure reforms, a process that is controlled by monitoring the viscoelastic moduli $G'$ and $G''$ at $f=1$~Hz and $\sigma=0.2$~Pa. A rest time of 100~s is sufficient for $G'$ and $G''$ to reach a steady state with negligible aging (Fig.~\ref{fig1}d). All measurements were performed at a temperature of 25$^\circ$C.

\subsection{Ultrasonic echography under shear}
\begin{figure}
\centering\includegraphics  [width=240pt] {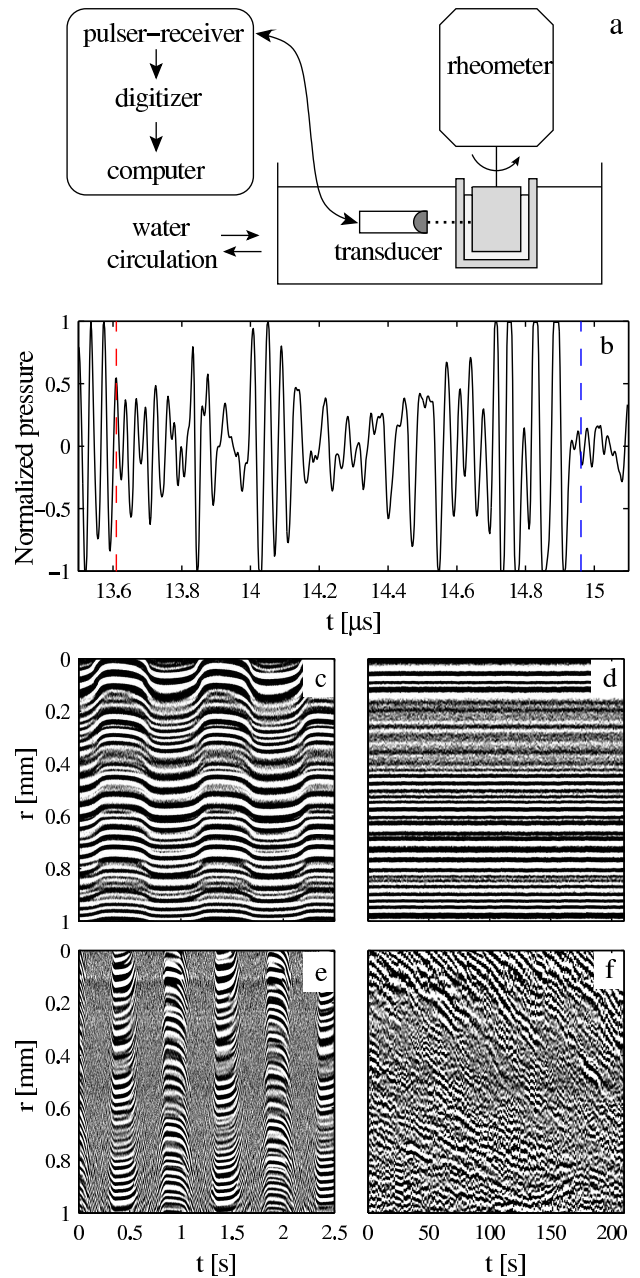}
\caption{Methods: ultrasonic echography under shear.
(a) Schematic of the rheology experiment combined with USV. Temperature is controlled to 25$\pm0.1^{\circ}$C by a water circulation around the Couette cell. (b) Typical normalized pressure signal backscattered by glass microspheres embedded in a carbon black gel as a function of time after the incident pulse is emitted at $t=0$~s. The times-of-flight corresponding to the positions of the stator and the rotor are indicated by a red and blue dashed line respectively. The corresponding gap is 1~mm. (c and d) Spatiotemporal diagrams of the normalized pressure signal during an oscillation experiment at $f=1$~Hz and $\sigma=15$~Pa. The pressure is coded in linear grey levels and plotted against the time at which the incident pulse is sent ($x$-axis) and the radial distance $r$ to the rotor ($y$-axis). In (c), the pulse repetition frequency $f_{\rm prf}$ is 400~Hz, which is large enough to resolve the deformation due to the oscillations of the rotor, while in (d), $f_{\rm prf}=f=1$~Hz. (e and f). Same as in (c and d) but for a higher shear stress $\sigma=30$~Pa.}
\label{fig5}
\end{figure}

In our experiments, classical rheology in Couette geometry is combined to ultrasonic echography in order to access local properties of the gel such as its velocity or deformation fields. In creep experiments, we use ultrasonic speckle velocimetry (USV) as described in Ref.~\cite{Manneville:2004a} to record temporally-resolved velocity profiles accross the gap at about 15~mm from the cell bottom (Fig.~\ref{fig5}a). USV is based on the interaction between ultrasound and micronsized scatterers embedded in the gel (here, the glass microspheres that act as contrast agents). A high-frequency (36~MHz) immersion transducer first emits a short ultrasonic pulse and collects the pressure signal backscattered from the gel. The backscattered signal is an ultrasonic speckle that directly reflects the spatial distribution of the scatterers within the gel along the acoustic beam (Fig.~\ref{fig5}b). As the scatterers move with the gel, cross-correlating two successive speckle signals over small time windows gives access to the local velocity of the gel with a spatial resolution of about 40~$\mu$m. This original combination of rheology and ultrasound turns out to be very efficient to follow the onset of flow both spatially and temporally.

For oscillation experiments, we take advantage of ultrasonic echography in a new way by focusing on the spatiotemporal diagrams of the successive speckle signals (Fig.~\ref{fig5}c--f). Indeed, the spatial variation of the speckle intensity provides crucial information on the local deformation of the sample. When pulses are sent with a sufficiently high repetition frequency $f_{\rm prf}$ compared to the oscillation frequency $f$, tracking the local oscillatory motion of the gel becomes possible (Fig.~\ref{fig5}c and e). Here, we rather set $f_{\rm prf}=f$ so that the material is probed by ultrasound only once per oscillation period. Data analysis is then straightforward: if two successive speckle signals are fully correlated, then the gel is solid-like since it comes back exactly to the same position every period (Fig.~\ref{fig5}d). On the other hand, we interpret any significant degree of decorrelation in the speckle signal from one period to the other as the signature of fluid-like behaviour due to motions of the scatterers induced by local rearrangements inside the gel (Fig.~\ref{fig5}f). Note that such decorrelation in the fluidized state does not arise from flow irreversibility (since the Reynolds number is $Re\simeq 10^{-3}\ll 1$) but rather from sedimentation of carbon black clusters due to the density mismatch between the mineral oil and the carbon black particles. We checked that sedimentation of clusters as small as 10~$\mu$m is sufficient to produce significant decorrelation after one oscillation period. The spatiotemporal tracking of the transition from correlated speckle signals (i.e. solid-like state) to decorrelated speckle signals (i.e. fluidized state) is made quantitative by using a threshold for the correlation coefficient of successive signals over small time windows, thus identifying the boundary between solid and fluid regions as a function of time up to a spatial resolution of 40~$\mu$m (Fig.~\ref{fig3}d).

The fraction of fluidized sample $\phi$ at a given time $t$ is readily extracted from the red contour separating fluid and solid regions in Fig.~\ref{fig3}d. Jumps in $\phi(t)$ are interpreted as fast fluidization events and identified using a threshold on the second derivative of $\phi(t)$. Statistical analysis of the time intervals $\delta t$ between two successive jumps then yields the probability distribution function $P$ and the complementary cumulative distribution function $C$ of $\delta t$ (inset of Fig.~\ref{fig4}). $C$ is defined as $C(x)=\int_x^\infty P(y){\rm d}y$ and experimentally estimated by sorting the measured values $\{\delta t_j\}_{j=1\dots N}$ in ascending order and by plotting $1-j/N$ as a function of $\delta t_j$ \cite{Berg:2008}.

\section{Results}

\subsection{Standard rheology}
\begin{figure}
\centering\includegraphics [width=240pt] {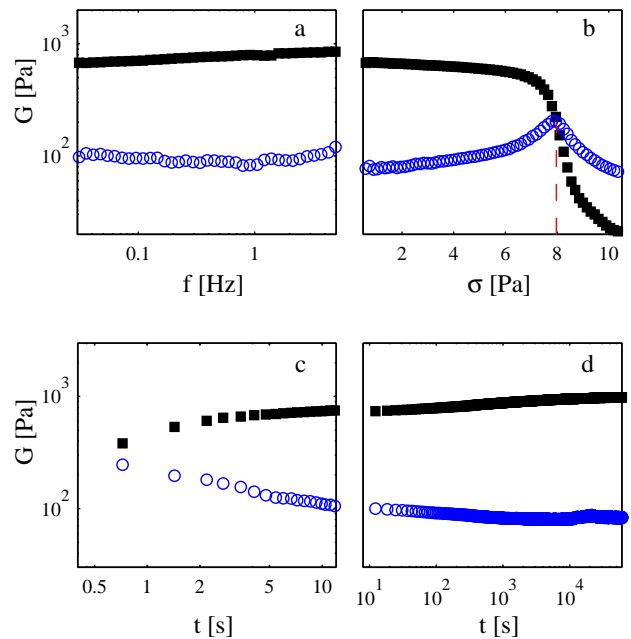}
\caption{Standard rheology of a 6~\% w/w carbon black suspension in a mineral oil. Viscoelastic moduli  $G'$ ($\blacksquare$) and $G''$ ({\color{blue}$\bigcirc$}) (a) in a frequency sweep in the linear regime at $\sigma=0.5$~Pa, (b) in a stress sweep experiment at $f=1$~Hz and (c, d) at $\sigma=0.2$~Pa and $f=3$~Hz as a function of time after preshear at high speed during 60~s. The red dashed line in (b) indicates the yield stress $\sigma_y$, which is time and frequency independent.}
\label{fig1}
\end{figure}

Our experiments focus on a colloidal gel made of carbon black particles dispersed at 6~\% w/w in a mineral oil \cite{Trappe:2000}. Its rheological properties are typical of a soft solid as shown in Fig.~\ref{fig1}: the elastic modulus $G'$ is about ten times larger than the loss modulus $G''$ and both moduli are frequency independent (Fig.~\ref{fig1}a). Moreover, as the amplitude $\sigma$ of the oscillatory shear stress is increased, the gel yields for $\sigma=\sigma_y\simeq 8$~Pa (Fig.~\ref{fig1}b), where $\sigma_y$ is defined by $G'(\sigma_y)=G''(\sigma_y)$. Above $\sigma_y$, one has $G''\gg G'$, i.e. the material behaves like a liquid. As seen in Fig.~\ref{fig1}c where the gel was fluidized by strong preshearing for times $t<0$ and left at rest for $t>0$, the elastic modulus recovers within a few seconds. This variation corresponds to the gelation process. The short range attraction between the carbon black particles leads to cluster aggregation. As the clusters percolate, they form a space-spanning network. The backbone of the network gives rise to the observed solid-like structure. For $t\gtrsim 100$~s, we observe very little aging (Fig.~\ref{fig1}d), which is ideal to run long experiments.

Yield stress measurements and global rheological properties are, however, too simple and elude fundamental issues such as the mechanism of fluidization and the role of self-healing processes due to the attractive interparticle forces. To follow the local dynamics of yielding, rheometry in Couette (concentric cylinders) geometry is coupled to ultrasonic speckle velocimetry (USV, see Methods section), which allows us to access deformation and velocity fields with high temporal resolution \cite{Manneville:2004a}, under either creep tests or oscillatory shear.

\subsection{Creep experiments}
\begin{figure}
\centering\includegraphics [width=240pt] {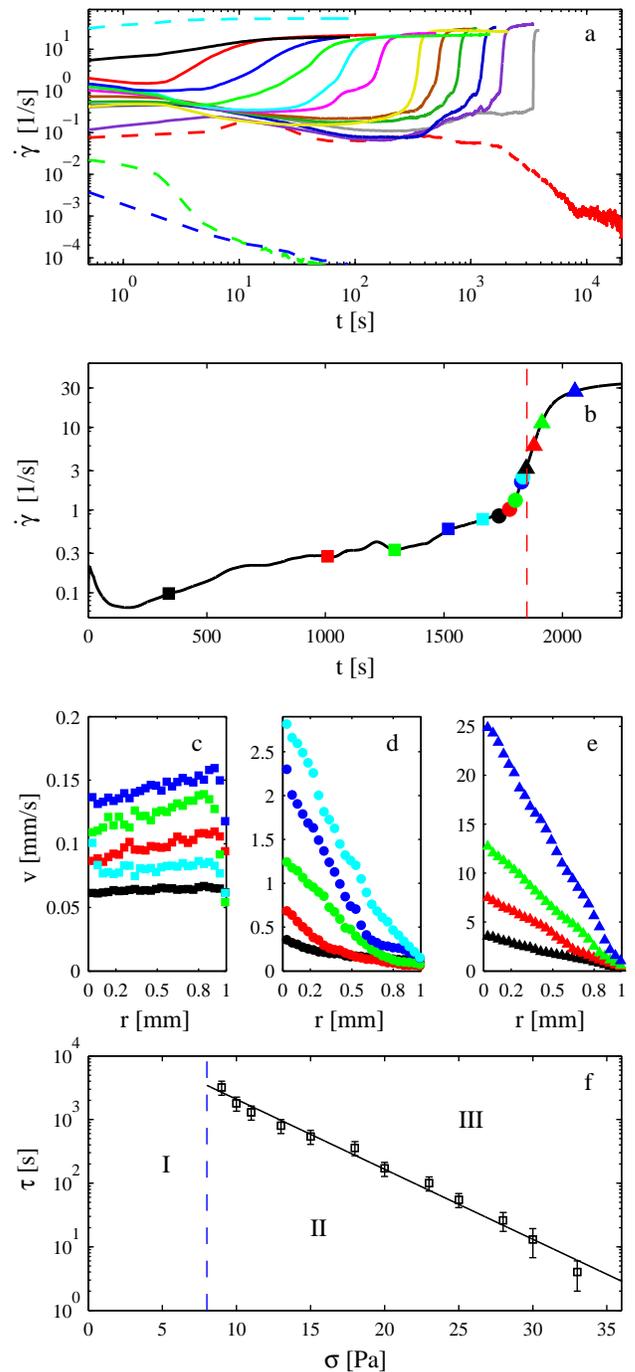}
\caption{Creep experiments in a 6~\% w/w carbon black gel.
(a) Temporal evolution of the shear rate $\dot{\gamma}$ for (from bottom to top) $\sigma=5$, 7, 8, 9, 10, 11, 13, 15, 18, 20, 23, 25, 28, 30, 33, and 40~Pa.
(b--e) Combined rheological and USV measurements for $\sigma=10$~Pa.
(b) $\dot{\gamma}(t)$ as a function of time. The red dashed line indicates the fluidization time $\tau$ discussed in the text. (c, d, e) Instantaneous velocity profiles $v(r)$ recorded with USV at the times shown in (b) using the same symbols.
(f) The fluidization time $\tau$ as a function of the applied shear stress $\sigma$. The black line is an exponential fit to the experimental data. Region I: solid-like behaviour. Region II: plug-like flow or shear localization. Region III: homogeneous flow.}
\label{fig2}
\end{figure}

Figure~\ref{fig2}a gathers the results of creep experiments for a wide range of imposed shear stresses. The critical shear stress $\sigma_y=8.5\pm 0.5$~Pa clearly separates two regimes. For $\sigma<\sigma_y$, the gel remains solid-like and does not flow, whereas for $\sigma>\sigma_y$, the sample eventually flows with a measurable shear rate $\dot{\gamma}$. The shear rate response $\dot{\gamma}(t)$ presents a characteristic shape that has already been observed in creep experiments on surfactant columnar phases  \cite{Bauer:2006}. In order to get better insight into the fluidization process, we correlate the rheological measurements to the local velocity field $v(r)$ measured with USV for the typical creep experiment at $\sigma=10$~Pa shown in Fig.~\ref{fig2}b. In the early stage where $\dot{\gamma}(t)$ slowly increases ($t\lesssim 1700$~s), solid-body rotation is observed, i.e. the sample is in an unsheared solid-like state except for thin lubrification layers along the cell walls that allow the gel to slide as a solid block (Fig.~\ref{fig2}c). At the beginning of the sharp upward bend in $\dot{\gamma}(t)$, for 1700~s~$\lesssim t \lesssim 1800$~s, the bulk sample starts to experience a non-zero shear rate. Fluidization then proceeds in a spatially heterogeneous way. Indeed, in this regime, velocity profiles present a highly sheared region close to the rotor that coexists with a low-shear band next to the stator and whose size increases in time (Fig.~\ref{fig2}d). Such shear localization suggests that the gel progressively sticks to the cell walls and gets dragged along the rotor. The fluidized zone expands from the rotor until, for $t\gtrsim 1800$~s, the gel flows homogeneously without any apparent wall slip, as seen from the linear velocity profiles of Fig.~\ref{fig2}e. In this last stage, the increase of the bulk shear rate indicates that further microstructural changes still occur and/or that sedimentation sets in.

Our local measurements refute the simple scenario in which yielding occurs abruptly and homogeneously at $\sigma_y$ (as suggested by the standard stress sweep experiment of Fig.~\ref{fig1}b) and rather support a yielding mechanism that is heterogeneous in both space and time, where fluidization starts at the walls and propagates through the whole sample. Such front dynamics is reminiscent of heterogeneous nucleation as predicted by a phenomenological ``fluidity'' model \cite{Picard:2002}. The present experiment also allows us to define unambiguously a characteristic time $\tau$ for fluidization as the time when homogeneous shear is first observed, e.g. $\tau=1850\pm 100$~s for $\sigma=10$~Pa. The fluidization time roughly corresponds to the inflection point in the $\dot{\gamma}(t)$ curve (and typically to $\dot{\gamma}=1$--10~s$^{-1}$), a criterion that has been used previously \cite{Bauer:2006} but without the confirmation from local measurements.

Finally, Fig.~\ref{fig2}f sums up the behavior of our colloidal gel in creep experiments. Below $\sigma_y$ (region I), the gel remains solid for all times, while above $\sigma_y$, the gel first undergoes solid body rotation or shear localization for $t<\tau$ (region II) and eventually flows homogeneously for $t>\tau$ (region III). Remarkably, the fluidization time follows an exponential law: $\tau\sim\exp(-\sigma/\sigma_0)$ where $\sigma_0=4.0\pm 0.2$~Pa. In other words, with $\sigma_0=k_B T/v$, where $v$ is some characteristic volume, this behaviour is equivalent to an Arrhenius law $\tau\sim\exp(E(\sigma)/(k_B T))$, where the energy barrier $E(\sigma)=-\sigma v$ decreases linearly with the applied stress.

\subsection{Oscillatory shear experiments}
\begin{figure}
\centering\includegraphics  [width=240pt] {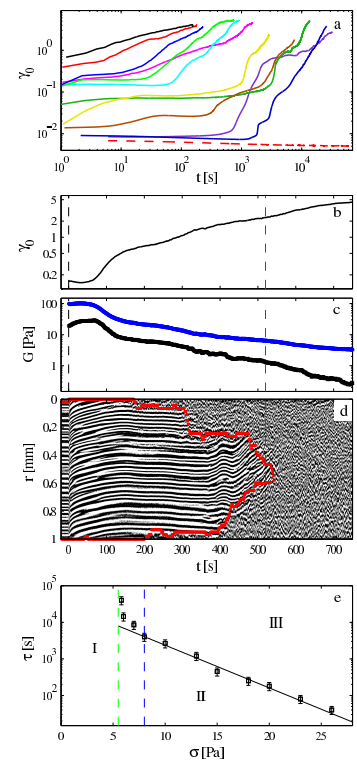}
\caption{Large amplitude oscillatory shear experiments at $f=1$~Hz in a 6~\% w/w carbon black gel.
(a) Temporal evolution of the deformation amplitude $\gamma_0$ for (from bottom to top) $\sigma=5$, 5.8, 6, 7, 8, 10, 13, 15, 18, 20, 23, 26~Pa.
(b-d) Combined rheological and USV measurements for $\sigma=15$~Pa.
(b) $\gamma_0$ and (c) $G'$ ({\color{black}$\bullet$}) and $G''$ ({\color{blue}$\bullet$}) as a function of time.
(d) Spatiotemporal diagram of the ultrasonic speckle signal coded in linear grey levels. The USV sampling frequency is equal to the oscillation frequency $f=1$~Hz. Red dots indicate the boundary between solid-like and fluid-like regions. The red dashed line shows the time $\tau$ at which 90~\% of the gel across the gap is fluidized.
(e) The fluidization time $\tau$ as a function of the stress amplitude $\sigma$.
The black line is an exponential fit to the experimental data (only data above 8~Pa are considered in the fit). Region I: solid-like behaviour. Region II: solid-fluid coexistence. Region III: fluid-like response. Between the green and blue dashed lines, $\tau$ strongly deviates from the exponential behaviour.}
\label{fig3}
\end{figure}

In a second series of experiments, the gel is submitted to oscillatory shear stress of given amplitude $\sigma$ and frequency $f=1$~Hz. The strain amplitude $\gamma_0(t)$ is monitored and again, we observe that when $\sigma$ is large enough $\gamma_0$ eventually increases with time, suggesting that the sample becomes fluid over time (Fig.~\ref{fig3}a,b). To confirm this scenario, we track the local deformation of the gel with USV (see Methods section). For $\sigma<5$~Pa, the gel indeed remains solid: no decorrelation of the ultrasonic speckle signals, i.e. no rearrangement within the gel, was observed during waiting times as long as $10^5$~s. However, for stress amplitudes above 5 Pa, rearrangements and fluid-like behaviour are detected first close to the cell walls (see $t\simeq 100$--300~s in Fig.~\ref{fig3}d). The fluidized zone then progressively expands through the bulk material ($t\simeq 300$--550~s) until the whole sample gets fluid ($t\gtrsim 550$~s).

It is important to note that fluidization does not have such a dramatic signature on $\gamma_0(t)$ under oscillatory stress as on $\dot{\gamma}(t)$ during creep tests. Moreover, as shown in Fig.~\ref{fig3}c, the loss modulus is larger than the elastic modulus throughout the whole experiment. In view of the USV results, this can only be explained by the presence of lubrication layers at the walls. A naive interpretation of standard viscoelastic data would thus wrongly conclude that the sample is fluid as soon as stress is applied. Here simultaneous spatially-resolved measurements show that solid--fluid coexistence comes into play and reveal a highly heterogeneous fluidization pattern.

In oscillation experiments, the fluidization time, defined as the time $\tau$ for which 90~\% of the gel is in the fluid-like state, follows an exponential law $\tau\sim\exp(-\sigma/\sigma_0)$ with $\sigma_0=3.9\pm 0.3$~Pa for $\sigma\gtrsim8$~Pa (Fig.~\ref{fig3}e). This is strikingly similar to the behaviour obtained in creep experiments (Fig.~\ref{fig2}f). However, under an oscillatory stress of amplitude 5~Pa~$\lesssim\sigma\lesssim8$~Pa, a strong departure from exponential behaviour is observed, pointing to a divergence at $\sigma\simeq 5.5$~Pa. In this stress range, no fluidization was observed during creep tests. Although this regime certainly deserves more attention, in the following, we shall focus on the exponential behaviour observed for stresses above 8~Pa.

\begin{figure}
\centering\includegraphics  [width=240pt] {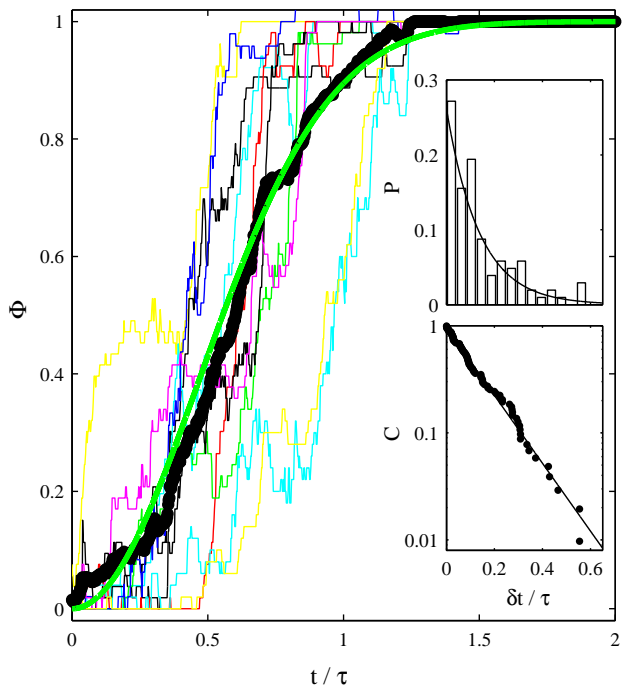}
\caption{Temporal evolution of the fluidized sample fraction $\phi$ for ten different oscillatory shear experiments at $f=1$~Hz and $\sigma=15$~Pa. Black dots show the average behaviour $\langle\phi(t)\rangle$ and the thick green line is $\langle\phi(t)\rangle=1-\exp(-t^2/\tilde{\tau}^2)$ with $\tilde{\tau}=320$~s. The data are plotted against $t/\tau$ where $\tau=480$~s is the fluidization time defined as $\langle\phi(\tau)\rangle=0.9$. Insets: probability distribution function $P$ (top) and complementary cumulative distribution function $C$ (bottom) of the waiting times $\delta t$ between two successive jumps in $\phi(t)$. The solid lines are exponential fits $P(\delta t)\sim C(\delta t)\sim\exp(-t/\tau^\star)$ with $\tau^\star = 0.137\tau=65.8$~s.}
\label{fig4}
\end{figure}

Shown in Fig.~\ref{fig4} is the fraction of fluidized sample $\phi(t)$ for ten different oscillatory experiments at a fixed stress amplitude of $\sigma=15$~Pa. The large deviations from one run to the other reveal the random nature of the local processes at work during yielding. These noisy features are compatible with models that include activated dynamics \cite{Kobelev:2005c,Fielding:2009,Sollich:1997}. Overall, the fluidized fraction $\langle\phi(t)\rangle$ averaged over the various experiments is well fitted by a sigmoidal relaxation $\langle\phi\rangle(t)=1-\exp(-t^2/\tilde{\tau}^2)$, a behaviour typically found in nucleation and one-dimensional growth phenomena \cite{Berret:1994a}. 

\section{Discussion}

The creep tests and the oscillation experiments above have shown the benefit of local measurements in establishing a spatiotemporal scenario for yielding: fluidization is initiated at the walls and propagates into the entire sample within a characteristic time $\tau$ that follows an Arrhenius law with an energy barrier $E=-\sigma v$. It is interesting to notice that a similar Arrhenius scaling has been reported in shear-induced aggregation of non-Brownian suspensions \cite{Guery:2006}. In both cases of shear-induced fluidization and aggregation, the energy barrier is reduced proportionally to the shear stress.

To the best of our knowledge, the present results provide the first experimental evidence for the relevance of activated processes and barrier hopping in the context of yielding in colloidal gels. Although such ingredients have recently been incorporated into theoretical approaches based on mode-coupling theory \cite{Kobelev:2005c}, local dynamical descriptions are still missing from microscopic models of the stress-induced solid-to-fluid transition. Our results also suggest a comparison with fracture in heterogeneous materials. Indeed, the propagation of fracture fronts is known to involve activated processes attibuted to the nucleation of microcracks \cite{Bonn:1998b,Guarino:2002,Tabuteau:2009}. Similarly, yielding proceeds by rapid local fluidization events (Fig.~\ref{fig3}d) separated by some ``waiting times'' $\delta t$. Previous measurements on fibrous materials have shown that time intervals between microcrack events are power-law distributed \cite{Guarino:2002}, a feature typical of self-similar avalanche-like phenomena. Here, $\delta t$ rather follows an exponential distribution (insets of Fig.~\ref{fig4}), which points to random uncorrelated events. Moreover, our observation of a linear energy barrier $E=-\sigma v$ is in stark contrast with the Griffith scaling $E\sim \sigma^{-4}$ for fracture \cite{Guarino:2002,Tabuteau:2009}. Following Ref.~\cite{Guery:2006}, the linear dependence $E=-\sigma v$ may be interpreted as the reduction of the energy barrier through hydrodynamic energy. In that case, the volume $v$ reads $v=a^2 \delta$ with $a$ the particle radius and $\delta$ a microscopic length characteristic of the interparticle potential. Here, assuming that $T$ can be identified with the thermodynamic temperature, $k_BT/v\simeq 4$~Pa and $a\simeq 500$~nm lead to $\delta\simeq 4$~nm, which provides a reasonable estimate for the range of attractive forces between carbon black particles.

Very recently, a novel rheological model to describe the local dynamics of ductile to brittle fracture was introduced and applied in the context of metallic glass formers \cite{Furukawa:2009}, but the connection with soft gels remains to be made. In particular, it can be noted that in some cases the fluidized fraction $\phi(t)$ decreases over short time periods before increasing again at longer times (Fig.~\ref{fig4}). This behaviour, absent from fracture phenomena, is a direct illustration of the competition between shear-activated cluster breaking (which favors fluidization and an increase of $\phi(t)$) and self-healing processes (which tend to lower $\phi(t)$) due to the attractive interactions.


In summary, the fluidization scenario depicted in this article highlights the crucial importance of time-resolved local measurements in characterizing the behaviour of soft materials under stress. Ultrasonic echography constitutes a versatile tool to follow yielding and fluidization in both space and time even in optically opaque materials. Its mesoscopic resolution of a few 10~$\mu$m bridges the gap between microscopic imaging techniques and global rheological measurements. Indeed, coarse-grained models for soft glassy materials use local strains and stresses averaged over intermediate sizes \cite{Bocquet:2009,Fielding:2009,Sollich:1997}, so that gathering information at a mesoscopic scale is a key issue for comparing experiments to such models. From the fundamental point of view, our observations not only emphasize the importance of activated processes and noise in yielding but also reveal a complex interplay between boundaries and bulk dynamics. Such results should provide a new test ground for computer simulations and for theoretical calculations in the current attempt to understand yielding in a wide variety of soft glassy materials.

\begin{acknowledgments}
The authors are grateful to L.~Bocquet, T.~Divoux, J.-F.~Palierne, and V.~Trappe for fruitful discussions.
\end{acknowledgments}


\begin{thebibliography}{10}

\bibitem{Larson:1999}
R.~G. Larson.
\newblock {\em The Structure and Rheology of Complex Fluids}.
\newblock Oxford University Press, 1999.

\bibitem{Russel:1989}
W.~B. Russel, D.~A. Saville, and W.~R. Schowalter.
\newblock {\em Colloidal Dispersions}.
\newblock Cambridge University Press (New York), 1989.

\bibitem{Cipelletti:2000}
L.~Cipelletti, S.~Manley, R.~C. Ball, and D.~A. Weitz.
\newblock Universal aging features in the restructuring of fractal colloidal
  gels.
\newblock {\em Phys. Rev. Lett.}, 84:2275--2278, 2000.

\bibitem{Duri:2006}
A.~Duri and L.~Cipelletti.
\newblock Length scale dependence of dynamical heterogeneity in a colloidal
  fractal gel.
\newblock {\em Europhys. Lett.}, 76:972--978, 2006.

\bibitem{Trappe:2007}
V.~Trappe, E.~Pitard, L.~Ramos, A.~Robert, H.~Bissig, and L.~Cipelletti.
\newblock Investigation of q-dependent dynamical heterogeneity in a colloidal
  gel by x-ray photon correlation spectroscopy.
\newblock {\em Phys. Rev. E}, 76:051404, 2007.

\bibitem{Trappe:2001}
V.~Trappe, V.~Prasad, L.~Cipelletti, P.~N. Segre, and D.~A. Weitz.
\newblock Jamming phase diagram for attractive particles.
\newblock {\em Nature}, 411:772--775, 2001.

\bibitem{Cardinaux:2007}
F.~Cardinaux, T.~Gibaud, A.~Stradner, and P.~Schurtenberger.
\newblock Interplay between spinodal decomposition and glass formation in
  proteins exhibiting short-range attractions.
\newblock {\em Phys. Rev. Lett.}, 99:118301, 2007.

\bibitem{Gao:2007}
Y.~Gao and M.~L. Kilfoil.
\newblock Direct imaging of dynamical heterogeneities near the colloid-gel
  transition.
\newblock {\em Phys. Rev. Lett.}, 99:078301, 2007.

\bibitem{Lu:2008}
P.~Lu, E.~Zaccarelli, F.~Ciulla, A.~B. Schofield, F.~Sciortino, and D.~A.
  Weitz.
\newblock Gelation of particles with short-range attraction.
\newblock {\em Nature}, 453:499--503, 2008.

\bibitem{Royall:2008}
C.~P. Royall, S.~R. Williams, T.~Ohtsuka, and H.~Tanaka.
\newblock Direct observation of a local structural mechanism for dynamic
  arrest.
\newblock {\em Nature Mater.}, 7:556--561, 2008.

\bibitem{Zaccarelli:2009}
E.~Zaccarelli and W.~C.~K. Poon.
\newblock Colloidal glasses and gels: The interplay of bonding.
\newblock {\em Proc. Natl. Acad. Sci. USA}, 106:15203--15208, 2009.

\bibitem{OHern:2003}
C.~S. O'Hern, L.~E. Silbert, A.~J. Liu, and S.~R. Nagel.
\newblock Jamming at zero temperature and zero applied stress: The epitome of
  disorder.
\newblock {\em Phys. Rev. E}, 68:011306, 2003.

\bibitem{Barnes:1999}
H.~A. Barnes.
\newblock The yield stress --- a review of 'panta rei' -- everything flows?
\newblock {\em J. Non-Newtonian Fluid Mech.}, 81:133--178, 1999.

\bibitem{Brader:2009}
J.~M. Brader, T.~Voigtmann, M.~Fuchs, R.~G. Larson, and M.~E. Cates.
\newblock Glass rheology: From mode-coupling theory to a dynamical yield
  criterion.
\newblock {\em Proc. Natl. Acad. Sci. USA}, 106:15186--15191, 2009.

\bibitem{Fuchs:2002}
M.~Fuchs and M.~E. Cates.
\newblock Theory of nonlinear rheology and yielding of dense colloidal
  suspensions.
\newblock {\em Phys. Rev. Lett.}, 89:248304, 2002.

\bibitem{Kobelev:2005c}
V.~Kobelev and K.~S. Schweizer.
\newblock Strain softening, yielding, and shear thinning in glassy colloidal
  suspensions.
\newblock {\em Phys. Rev. E}, 71:021401, 2005.

\bibitem{Osuji:2008}
C.~O. Osuji, C.~Kim, and D.~A. Weitz.
\newblock Shear thickening and scaling of the elastic modulus in a fractal
  colloidal system with attractive interactions.
\newblock {\em Phys. Rev. E}, 77:060402(R), 2008.

\bibitem{Potanin:1995}
A.~A. Potanin, R.~De Rooij, D.~Van den Ende, and J.~Mellema.
\newblock Microrheological modeling of weakly aggregated dispersions.
\newblock {\em J. Chem. Phys.}, 102:5845--5853, 1995.

\bibitem{Rueb:1997}
C.~J. Rueb and C.~F. Zukoski.
\newblock Viscoelastic properties of colloidal gels.
\newblock {\em J. Rheol.}, 41:197--218, 1997.

\bibitem{Whittle:1998}
M.~Whittle and E.~Dickinson.
\newblock Large deformation rheological behaviour of a model particle gel.
\newblock {\em J. Chem. Soc., Faraday Trans.}, 94:2453--2462, 1998.

\bibitem{Woutersen:1990}
A.~T. J.~M. Woutersen and C.~G. de~Kruif.
\newblock The rheology of adhesive hard sphere dispersions.
\newblock {\em J. Chem. Phys.}, 94:5739--5750, 1990.

\bibitem{Barnes:1995}
H.~A. Barnes.
\newblock A review of the slip (wall depletion) of polymer solutions, emulsions
  and particle suspensions in viscometers: Its cause, character, and cure.
\newblock {\em J. Non-Newtonian Fluid Mech.}, 56:221--251, 1995.

\bibitem{Coussot:2002a}
P.~Coussot, J.~S. Raynaud, F.~Bertrand, P.~Moucheront, J.~P. Guilbaud, H.~T.
  Huynh, S.~Jarny, and D.~Lesueur.
\newblock Coexistence of liquid and solid phases in flowing soft-glassy
  materials.
\newblock {\em Phys. Rev. Lett.}, 88:218301, 2002.

\bibitem{Pignon:1996}
F.~Pignon, A.~Magnin, and J.-M. Piau.
\newblock Thixotropic colloidal suspensions and flow curves with minimum:
  Identification of flow regimes and rheometric consequences.
\newblock {\em J. Rheol.}, 40:573--587, 1996.

\bibitem{Ballesta:2008b}
P.~Ballesta, R.~Besseling, L.~Isa, G.~Petekidis, and W.~C.~K. Poon.
\newblock Slip and flow of hard-sphere colloidal glasses.
\newblock {\em Phys. Rev. Lett.}, 101:258301, 2008.

\bibitem{Besseling:2007}
R.~Besseling, E.~R. Weeks, A.~B. Schofield, and W.~C.~K. Poon.
\newblock Three-dimensional imaging of colloidal glasses under steady shear.
\newblock {\em Phys. Rev. Lett.}, 99:028301, 2007.

\bibitem{Isa:2007}
L.~Isa, R.~Besseling, and W.~C.~K. Poon.
\newblock Shear zones and wall slip in the capillary flow of concentrated
  colloidal suspensions.
\newblock {\em Phys. Rev. Lett.}, 98:198305, 2007.

\bibitem{Goyon:2008}
J.~Goyon, A.~Colin, G.~Ovarlez, A.~Ajdari, and L.~Bocquet.
\newblock Flow cooperativity and breakdown of local constitutive laws for
  confined glassy flows.
\newblock {\em Nature}, 454:84--87, 2008.

\bibitem{Bocquet:2009}
L.~Bocquet, A.~Colin, and A.~Ajdari.
\newblock A kinetic theory of plastic flow in soft glassy materials.
\newblock {\em Phys. Rev. Lett.}, 103:036001, 2009.

\bibitem{Fielding:2009}
S.~M. Fielding, M.~E. Cates, and P.~Sollich.
\newblock Shear banding, aging and noise dynamics in soft glassy materials.
\newblock {\em Soft Matter}, 5:2378--2382, 2009.

\bibitem{Donnet:1993}
J.-B. Donnet, R.~C. Bansal, and M.-J. Wang.
\newblock {\em Carbon black: Science and technology}.
\newblock Marcel Dekker Inc. (New York), 1993.

\bibitem{Manneville:2004a}
S.~Manneville, L.~B{\'e}cu, and A.~Colin.
\newblock High-frequency ultrasonic speckle velocimetry in sheared complex
  fluids.
\newblock {\em Eur. Phys. J. AP}, 28:361--373, 2004.

\bibitem{Berg:2008}
B.~Berg and R.~Harris.
\newblock From data to probability densities without histograms.
\newblock {\em Comput. Phys. Commun.}, 179:443--448, 1999.

\bibitem{Trappe:2000}
V.~Trappe and D.~A. Weitz.
\newblock Scaling of the viscoelasticity of weakly attractive particles.
\newblock {\em Phys. Rev. Lett.}, 85:449--452, 2000.

\bibitem{Bauer:2006}
T.~Bauer, J.~Oberdisse, and L.~Ramos.
\newblock Collective rearrangement at the onset of flow of a polycrystalline
  hexagonal columnar phase.
\newblock {\em Phys. Rev. Lett.}, 97:258303, 2006.

\bibitem{Picard:2002}
G.~Picard, A.~Ajdari, L.~Bocquet, and F.~Lequeux.
\newblock A simple model for heterogeneous flows of yield stress fluids.
\newblock {\em Phys. Rev. E}, 66:051501, 2002.

\bibitem{Sollich:1997}
P.~Sollich, F.~Lequeux, P.~H\'ebraud, and M.~E. Cates.
\newblock Rheology of soft glassy materials.
\newblock {\em Phys. Rev. Lett.}, 78:2020--2023, 1997.

\bibitem{Berret:1994a}
J.-F. Berret, D.~C. Roux, and G.~Porte.
\newblock Isotropic-to-nematic transition in wormlike micelles under shear.
\newblock {\em J. Phys. II France}, 4:1261--1279, 1994.

\bibitem{Guery:2006}
J.~Guery, E.~Bertrand, C.~Rouzeau, P.~Levitz, D.~A. Weitz, and J.~Bibette.
\newblock Irreversible shear-activated aggregation in non-brownian suspensions.
\newblock {\em Phys. Rev. Lett.}, 96:198301, 2006.

\bibitem{Bonn:1998b}
D.~Bonn, H.~Kellay, M.~Prochnow, K.~Ben-Djemiaa, and J.~Meunier.
\newblock Delayed fracture of an inhomogeneous soft solid.
\newblock {\em Science}, 280:265--267, 1998.

\bibitem{Guarino:2002}
A.~Guarino, S.~Ciliberto, A.~Garcimart{\'\i}n, M.~Zei, and R.~Scorretti.
\newblock Failure time and critical behaviour of fracture precursors in
  heterogeneous materials.
\newblock {\em Eur. Phys. J. B}, 26:141--151, 2002.

\bibitem{Tabuteau:2009}
H.~Tabuteau, S.~Mora, G.~Porte, M.~Abkarian, and C.~Ligoure.
\newblock Microscopic mechanisms of the brittleness of viscoelastic fluids.
\newblock {\em Phys. Rev. Lett.}, 102:155501, 2009.

\bibitem{Furukawa:2009}
A.~Furukawa and H.~Tanaka.
\newblock Inhomogeneous flow and fracture of glassy materials.
\newblock {\em Nature Mater.}, 8:601--609, 2009.

\end{thebibliography}

\end{document}